\documentclass[a4paper,12pt]{article}

\usepackage[latin1]{inputenc}
\usepackage[top=2.5cm, left=2.5cm, right=2cm, bottom=2.5cm]{geometry}
\usepackage{graphicx}
\usepackage[centertags]{amsmath}
\usepackage{amsfonts}
\usepackage{amssymb}
\usepackage{amsthm}
\usepackage{newlfont}

\begin{document}

\begin{center}

{\Large Three-dimensional Background Field Gravity: A Hamilton-Jacobi analysis}

\vspace{1cm}

N. T. Maia\footnote{nmaia@ift.unesp.br}, B. M. Pimentel\footnote{pimentel@ift.unesp.br}, C. E. Valc\'arcel\footnote{valcarcel.flores@gmail.com}

\vspace{.5cm}

$^{1,2}$\emph{Instituto de F\'isica Te\'orica, UNESP - S\~ao Paulo State University},

\emph{P. O. Box 70532-2, 01156-970, S\~ao Paulo, SP, Brazil}.

$^{3}$\emph{Centro de Matem\'atica, Computa\c{c}\~ao e Cogni\c{c}\~ao},

\emph{Universidade Federal do ABC, 09210-170 Santo Andr\'e, SP, Brazil}.

\end{center}

\vspace{.25cm}

\begin{abstract}

We analyse the constraint structure of the Background Field model for three dimensional
gravity including a cosmological term via the Hamilton-Jacobi formalism.
We find the complete set of involutive Hamiltonians that assures the integrability of
the system and calculate the characteristic equations of the system. We established the
equivalence between these equations and the field equations and also obtain the generators
of canonical and gauge transformations.\\

\vspace{.5cm}


\end{abstract}

\section{Introduction}

Topological Quantum Field Theories (TQFT) were introduced by Witten \cite{Witten} at the late 80s
and until now they have found a wide range of applications in Physics. One characteristic of
these theories is that their correlation functions do not depend on the space-time metric. According to
Birmingham \cite{Birmingham} the TQFT can be divided into two groups: the Witten (or cohomological) type
and the Schwarz type. The Chern-Simons (CS) gauge theory is a Schwarz type TQFT defined in odd dimensions
which is used, for example, in addition with three-dimensional kinetic actions to build the so called Topologically
Massive Theories \cite{Deser}.

The Background Field (BF) model is another Schwarz type TQFT and had been widely used due to its relation
with Gravity. For example, there has been shown that the two-dimensional BF model can be equivalent to the two-dimensional
Jackiw-Teitelboim Gravity \cite{JT} for a given gauge group \cite{JTBF}. The three-dimensional BF model
is equivalent to the first order formulation of pure General Relativity under the Lorentz gauge group $SO(2,1)$ \cite{Oda}
and the four-dimensional gravity is equivalent to the Plebanski action\cite{Plebanski} which consists in a BF action
plus a Lagrangian multiplier. An extensive review between these equivalences can be found in \cite{Freidel}.

The BF lower dimensional models of gravity are good laboratories for the study of spin foam quantization \cite{Perez}
and loop quantum gravity. In both schemes of quantization, the simplectic structure of the BF model is of utmost importance.
In order to identify the correct phase space, the Dirac canonical analysis \cite{Dirac} is one of the most used tools. This analysis has
been done in two \cite{Piguet} and three \cite{Escalante} dimensional BF models of gravity. Nonetheless, there
are other schemes of constraint analysis, as the Faddeev-Jackiw \cite{FJ} formalism and the Hamilton-Jacobi (HJ) formalism.

A first attempt to use the HJ formalism as an approach to constrained systems was given by Dominici, et. al. \cite{Pons}. Here we will deal
with the approach developed by G\"{u}ler \cite{Guler1} as an extension to the Carath\'{e}odory's equivalent Lagrangians method to the
calculus of variations \cite{Caratheodory}. The conditions for stationary action are reduced to a set of Hamilton-Jacobi partial
differential equations, also called Hamiltonians, that must obey the Frobenius' Integrability Condition (IC). In \cite{Pimentel-NONINV} has been shown that
in order to satisfy the IC the non-involutive Hamiltonians must be eliminated, this way they redefine the dynamic of the system by
building the Generalised Bracket (GB). Therefore, we end with a set of complete involutive Hamiltonians, which plays the role of generators of the
canonical transformations  \cite{Pimentel-INV}. The Hamilton-Jacobi formalism has been generalised to higher order Lagrangians and Berezin systems,
among others \cite{HJ-Generalizations}, as well as applied to different kind of physical systems, more recently to Topologically Massive
theories \cite{Pimentel-TMYM} and gravity models \cite{HJ-Gravity}, including the two-dimensional BF gravity \cite{Pimentel-BF}. In this article we
will apply the HJ formalism to the three-dimensional BF model for gravity.

In the following section we will shown the HJ formalism (for a more detailed explanation see \cite{Pimentel-NONINV}\cite{Pimentel-INV}). In
section $3$ the three-dimensional BF gravity will be presented. In section $4$ we will perform its Hamilton-Jacobi constraint analysis and build the
Generalised Brackets. In section $5$ we will compute the characteristic equations (CE) and analyse the dynamical evolution along the independent
parameters of the theory. From this analysis we obtain the equivalence between the Lagrangian equations of motion and the temporal evolution
of the CE. From the evolution along the parameters related the involutive Hamiltonians, we obtain the generators of canonical and gauge transformations.
In section $6$ we will discuss the results.

\bigskip

\section{The Hamilton-Jacobi Formalism}

Let us consider a physical system with a Lagrangian function $L=L(x^{i},\dot{x}%
^{i},t)$, where the Latin indices $i,j$ go from $1$ to $n$, being $n$ the dimension
of the configuration space. This Lagrangian is called singular or constrained if it does not
satisfy the Hessian Condition, which states that the matrix elements $W _{ij}=\frac{\partial ^{2}L}{\partial \dot{x}^{i}\partial \dot{x}^{j}}$
has a determinant equal to zero. Whenever the Hessian Condition ($\det W_{ij} \neq 0$) is not satisfied, it is
implied that some of the conjugated momenta $p_i=\frac{\partial L}{\partial \dot{x}^i}$ are not
invertible on the velocities.  By considering $k$ non-invertible momenta and $m=n-k$ invertible
momenta, we have
\begin{equation}
p_{z}-\frac{\partial L}{\partial \dot{x}^{z}}=0,
\end{equation}
where $z=1,...,k$. Defining $H_{z}\equiv -\frac{\partial L}{\partial \dot{x}%
^{z}}$, the above equation is rewritten as
\begin{equation}
H_{z}^{\prime }\equiv p_{z}+H_{z}=0.  \label{ham1}
\end{equation}
We call \textit{Hamiltonians} the constraints represented in this way.
Defining $p_{0}\equiv \frac{\partial S}{\partial t}$ , the HJ equation is
the Hamiltonian
\begin{equation}
H_{0}^{\prime }\equiv p_{0}+H_{0}=0.  \label{ham2}
\end{equation}

The canonical Hamiltonian function $H_{0}=p_{a}\dot{x}^{a}+p_{z}\dot{x}%
^{z}-L $, regarding $a=1,...,m$, does not depend on the non-invertible
velocities $\dot{x}^{z}$ if the the constraints are carried out. Putting
together (\ref{ham1}) and (\ref{ham2}), we form the initial set of
Hamilton-Jacobi Partial Differential Equations (HJPDE):

\begin{equation}
H_{\alpha }^{\prime }\equiv p_{\alpha }+H_{\alpha }=0,  \label{hamtot}
\end{equation}

where $\alpha =0,1,...,k$. Through the Cauchy's method \cite{Caratheodory}, the characteristic
equations related to the first order equations system (\ref{hamtot}) are
given by

\begin{equation}
dx^{a}=\frac{\partial H_{\alpha }^{\prime }}{\partial p_{a}}dt^{\alpha
},\qquad dp_{a}=-\frac{\partial H_{\alpha }^{\prime }}{\partial x^{a}}%
dt^{\alpha },\qquad dS=\left( p_{a}dx^{a}-H_{\alpha }dt^{\alpha }\right) .
\label{eqcar}
\end{equation}

From these differential equations, the Poisson Brackets (PB) defined on the
extended phase space ($x^{a},t^{\alpha },p_{a},p_{\alpha }$) can be used to
express in a concise form the evolution of any function $f=f(x^{a},t^{\alpha
},p_{a},p_{\alpha })$:

\begin{equation}
df=\left\{ f,H_{\alpha }^{\prime }\right\} dt^{\alpha }.  \label{ev}
\end{equation}

This is the \textit{fundamental differential} whereby the Hamiltonians can
be seen as the generators of the dynamical evolution of the phase space
functions.

A geometrical interpretation can be given at this point. The solutions of
the first two equations of (\ref{eqcar}) give rise to a congruence of curves
on the reduced phase space ($x^{a},p_{a}$). The characteristic curves $%
x^{a}(t,x^{z})$ describe the dynamical trajectories and depend on the $k+1$
parameters $t^{\alpha }$ which in turn must be regarded as the \textit{%
independent variables} of the system. A complete solution of (\ref{hamtot})
is given by a family of surfaces orthogonal to the characteristic curves and
its existence is ensured by satisfying the Frobenius' integrability condition \cite{Pimentel-INV}
which is written as

\begin{equation}
\left\{ H_{\alpha }^{\prime },H_{\beta }^{\prime }\right\} =C_{\alpha \beta
}^{\rule{9pt}{0pt}\gamma }H_{\gamma }^{\prime }.  \label{c1}
\end{equation}
It means the Hamiltonians must close a \textit{Lie algebra}. Equivalently,
\begin{equation}
dH_{\alpha }^{\prime }=0.  \label{c2}
\end{equation}

Hamiltonians that satisfy the Frobenius integrability condition are called
\textit{involutives} while the \textit{non-involutives} are those that do
not satisfy it. We can add new constraints to the system imposing condition (%
\ref{c2}) and then completing the set of HJPDE. However, sometimes this
procedure is not sufficient to make the set of HJPDE integrable. When the
condition (\ref{c2}) is imposed, some Hamiltonians may provide relations
that exhibit dependence between some parameters. These Hamiltonians can be
used to construct a new algebra which we call the \textit{Generalised
Brackets} (GB):

\begin{equation*}
\left\{ A,B\right\} ^{\ast }=\left\{ A,B\right\} -\left\{ A,H_{\overline{a}%
}^{\prime }\right\} \left( M_{\overline{a}\overline{b}}\right) ^{-1}\left\{
H_{\overline{b}}^{\prime },B\right\} .
\end{equation*}

The indices $\overline{a}$ and $\overline{b}$ are related to the non-involutive
Hamiltonians whose parameters are somehow related. The matrix $M$ is built
from the PB of these Hamiltonians, i.e., its elements are $M_{\overline{a}%
\overline{b}}=\left\{ H_{\overline{a}}^{\prime },H_{\overline{b}}^{\prime
}\right\} $. In this way, these non-involutive Hamiltonians are absorbed in
the new algebra. The integrability of the remaining Hamiltonians must be
analysed through the GB algebra instead of the PB algebra. New Hamiltonians
may be added to complete the HJPDE set in this process until we get as
result an integrable set of HJPDE.

Let us define the variables on the extended phase space as $z^I=(x^a, t^\alpha, p_a,p_\alpha)$
and define the vector field $X_\alpha$ with components
\begin{equation}
X_{\alpha }^{I}\equiv \left\{ z^{I},H_{\alpha }^{\prime }\right\}^* ,\label{HJS35}
\end{equation}
such that, any function on the extended phase space can be written as
\begin{equation}
dF=\left\{ F,H_{\alpha }^{\prime }\right\}^* dt^{\alpha }=X_{\alpha }\left[ F%
\right] dt^{\alpha }.  \label{HJS38}
\end{equation}%
The vector $X_{\alpha }$ are related to the dynamical evolution of the system, since the
CE are included on (\ref{HJS38}). From the definition of vectors $X_{\alpha }$ and using the Jacobi Identity
we obtain
\begin{equation}
\left[ X_{\alpha },X_{\beta }\right] F=\left\{ \left\{ F,H_{\beta }^{\prime
}\right\}^* ,H_{\alpha }^{\prime }\right\}^* -\left\{ \left\{ F,H_{\alpha
}^{\prime }\right\}^* ,H_{\beta }^{\prime }\right\}^* =\left\{ \left\{ H_{\alpha
}^{\prime },H_{\beta }^{\prime }\right\}^* ,F\right\}^* ,
\end{equation}%
Whenever the system is integrable, i.e., (\ref{c1}) or (\ref{c2}) are valid, we can write
\begin{equation}
\left[ X_{\alpha },X_{\beta }\right] F=-\left\{ f_{\ \alpha \beta }^{\gamma
}H_{\gamma }^{\prime },F\right\}^* = f_{\ \alpha \beta }^{\gamma }X_{\gamma }%
\left[ F\right] - \left\{ f_{\ \alpha \beta }^{\gamma },F\right\}^* H_{\gamma
}^{\prime },
\end{equation}
where $f_{\ \alpha \beta }^{\gamma}=-C_{\ \alpha \beta }^{\gamma}$. If the structure constants are independent of the variables
of the extended phase space the IC becomes a condition over the commutator
\begin{equation}
\left[ X_{\alpha },X_{\beta }\right] = f_{\ \alpha \beta }^{\gamma
}X_{\gamma },  \label{HJS41}
\end{equation}%
which is, indeed, the necessary condition for $X_\alpha$ to be a complete basis.

In general, a transformation of a function $F$ can be written as
\begin{equation}
\delta F=\delta t^{\alpha }X_{\alpha }F,  \label{HJS50}
\end{equation}%
where $\delta t^{\alpha}=\bar{t}^{\alpha }-t^{\alpha }$ are arbitrary functions of $z^{I}$. However, notice that if we choose $\delta t^{\alpha }=dt^{\alpha }$,
equation (\ref{HJS50}) becomes the fundamental differential. For any variable of the extended phase space $z^{I}$
we have
\begin{equation}
\delta z^{I}=\bar{z}^{I}\left( \bar{t}^{\alpha }\right) -z^{I}\left(
t^{\alpha }\right) =\delta t^{\alpha }X_{\alpha }\left[ z^{I}\right] .
\label{HJS51}
\end{equation}%
Now, let us consider a transformation $g$ such that
\begin{equation}
\bar{z}^{I}\left( \bar{t}^{\alpha }\right) =gz^{I}\left( t^{\alpha }\right) .
\label{HJS52}
\end{equation}%
In this case
\begin{equation}
g=1+\delta t^{\alpha }X_{\alpha }.  \label{HJS53}
\end{equation}%
We say that transformation $g$ carries the infinitesimal flows generated by the vectors $X_\alpha$. This is what we
call characteristics flows (CF). It can be shown that whenever the IC is satisfied, the transformation $g$ has an inverse
\begin{equation}
g^{-1}=1-\delta t^{\alpha }X_{\alpha }.
\end{equation}%
and also preserve the symplectic structure $\omega \equiv dx^a \wedge dp_a + dt^\alpha \wedge dp_\alpha + dH_\alpha \wedge dt^\alpha$
\begin{equation}
g\omega g^{-1}=\omega.
\end{equation}
This show that $g$ are canonical transformations and that the complete set of involutive
Hamiltonians $H_{\alpha }^{\prime }$ are the generators of these transformations.

In order to relate the canonical transformations with the gauge ones, we need to restrict the study to fixed times
$\delta t^{0}=\delta t=0$, which is the classical equivalent to a fixed point transformation in field theory. The
transformation on any variable $z^{I}$ now reads
\begin{equation}
\delta z^{I}=\left\{ z^{I},H_{z}^{\prime }\right\}^* \delta t^{z},
\label{HJS55}
\end{equation}%
If we can keep this transformation canonical, the IC must be satisfied, this is
\begin{equation}
\left\{ H_{x}^{\prime },H_{y}^{\prime }\right\}^* =C_{\ xy}^{z}H_{z}^{\prime },
\label{HJS56}
\end{equation}%
Nonetheless, this condition does not guarantee the integrability on the algebra of the Hamiltonians, which is
\begin{equation}
\left\{ H_{x}^{\prime },H_{y}^{\prime }\right\}^* =C_{\ xy}^{0}H_{0}^{\prime
}+C_{\ xy}^{z}H_{z}^{\prime }.  \label{HJS57}
\end{equation}%
To conciliate both equations we must consider whether $C_{\ xy}^{0}=0$ or $H_{0}^{\prime }=0$. However, condition
$C_{\ xy}^{0}=0$ is too strong since it implies that $\left\{ H_{0}^{\prime
},H_{z}^{\prime }\right\} =0$, which is almost never satisfied. On the other hand, the condition $%
H_{0}^{\prime }=0$ constrains the phase space. Under this assumption, we define
\begin{equation}
G^{can}\equiv H_{z}^{\prime }\delta t^{z},  \label{HJS58}
\end{equation}%
which is the generator of the canonical transformations, once that
\begin{equation}
\delta z^{I}=\left\{ z^{I},G^{can}\right\}^* .  \label{HJS59}
\end{equation}

\bigskip

\section{Three-Dimensional BF model}

Let us consider a $d$-dimensional manifold $\mathcal M$, a Lie group $G$, a
connection $A$ and a $(d-2)$-form $B$ called Background Field. With those elements let us build the following action
\begin{equation}
W_{BF}=\int_{\mathcal M}tr[B\wedge F],\label{3DBF01}
\end{equation}
where $F$ is the curvature of the connection $A$, i.e., $F=DA$. Due to the properties of the trace and the
exterior product $\wedge$ it is straightforward to see that this action is gauge invariant.

In three dimensions we can add another
invariant $tr[B\wedge B\wedge B]$. Therefore, the three-dimensional BF action can be written as
\begin{equation}
W_{BF}=\int_{\mathcal M}tr\left(B\wedge F(A) + \kappa \, B\wedge B\wedge B\right) .\label{3DBF02}
\end{equation}
where $\kappa$ is a constant. Due to its construction, (\ref{3DBF02}) is invariant under gauge transformation:
\begin{equation}
\delta A = D \chi,\ \ \ \delta B = [B,\chi], \label{3DBF03}
\end{equation}
but also quasi-invariant under shift transformation:
\begin{equation}
\delta B = D \eta,\ \ \ \delta A = 3 \kappa [B,\eta], \label{3DBF04}
\end{equation}
being $\xi$ and $\eta$ arbitrary functions.

It has been shown that in three dimensions and, considering $G$ as the Lorentz group $SO(1,2)$, the
BF action (\ref{3DBF02}) is equivalent to Einstein-Hilbert-Palatini gravity in terms of vielbeins. Therefore, considering
$G$ as the Lorentz group and $\kappa=-\Lambda /3$, the action (\ref{3DBF02}) represents Riemann gravity plus
Cosmological constant.

Before proceed with any kind of quantization scheme, the reduced phase space of the system must be well
defined. The determination of the true degrees of freedom are determined after the analysis of the constrains of the theory.

\bigskip

\section{The Hamilton-Jacobi analysis of the 3D BF gravity}

The constraint analysis is not covariant. We refer to one specific time choice to build the HJ equations. It is, then, appropriated to
leave the differential forms notation and write the Lagrangian in terms of the components of the background and gauge field, i.e.,
\begin{equation}
A=A^a_\mu J_a dx^\mu, \, \, B=B^a_\mu J_a dx^\mu,
\end{equation}
where $J_a$ are generators of the $G=SO(1,2)$ group. These generators satisfy $[J_a,J_b]=f_{abc}J_c$ and $tr(J_aJ_b)=\frac{1}{2}\eta_{ab}$, where
$\eta_{ab}=diag(+,-,-)$ . Therefore
\begin{equation}
\mathcal{L}=\frac{1}{2}\epsilon ^{\mu \gamma \nu }(B_{a\mu }F_{\gamma \nu
}^{a} - \frac{\Lambda}{3} f_{abc}B_{\mu }^{a}B_{\gamma }^{b}B_{\nu }^{c}), \label{HJBF01}
\end{equation}
where $F^a_{\mu\nu} = \partial_\mu A^a_\nu - \partial_\nu A^a_\mu+ f^a_{\ bc}A^b_\mu A^c_\nu$. The equations of motion are
\begin{eqnarray}
0 &=& \epsilon ^{\mu \gamma \nu }\left( F_{\gamma \nu }^{a}-\Lambda f_{\rule{4pt}{0pt}%
bc}^{a}B_{\gamma }^{b}B_{\nu }^{c}\right),  \label{HJBF02}\\
0 &=& \epsilon ^{\mu \gamma \nu }D_{\gamma }B_{\nu }^{a}. \label{HJBF03}
\end{eqnarray}
Here we had made use of the definition of covariant derivative
\begin{equation}
D_{\mu }\theta _{\nu }^{a}\equiv \partial
_{\mu }\theta _{\nu }^{a}+f_{\ bc}^{a}A_{\mu }^{b}\theta _{\nu
}^{c}.\label{HJBF04}
\end{equation}
Furthermore, equation (\ref{HJBF02}) represent the dynamical equation of three-dimensional gravity, and (\ref{HJBF03}) represent
the zero torsion condition.

Now, to begin with the HJ analysis of the three-dimensional BF gravity, we compute the
momenta $\pi^a$ and $\Pi^a$ conjugated to $A^a_\mu$ and $B^a_\mu$ respectively
\begin{eqnarray}
\pi _{a}^{\mu } &\equiv& \frac{\partial \mathcal{L}}{\partial \partial
_{0}A_{\mu }^{a}}=\epsilon ^{0\mu\nu}B_{a\nu },\label{HJBF05}\\
\Pi _{a}^{\mu } &\equiv& \frac{\partial \mathcal{L}}{\partial \partial
_{0}B_{\mu }^{a}}=0.\label{HJBF06}
\end{eqnarray}
The expressions above do not depend on any velocities $\partial
_{0}A_{\mu }^{a}$, $\partial
_{0}B_{\mu }^{a}$. Therefore they are canonical constraints of the theory. It turns out the canonical Hamiltonian density is given by
\begin{equation}
\mathcal{H}_{0}= - \epsilon ^{0\gamma \nu } \left[ A_{a0}D_{\gamma }B_{\nu
}^{a} + B_{a0} \left( F_{\gamma\nu}^{a}-\Lambda f_{\
bc}^{a} B_{\gamma }^{b}B_{\nu }^{c} \right) \right]  .\label{HJBF07}
\end{equation}

Let us define $\pi \equiv \partial _{0}S$. Then, the initial set of HJPDE is
\begin{eqnarray}
\mathcal{H}^{\prime } &\equiv &\pi +\mathcal{H}_{0}=0, \\
\mathcal{A}_{a}^{\prime 0} &\equiv &\pi _{a}^{0}=0, \\
\mathcal{A}_{a}^{\prime 1} &\equiv &\pi _{a}^{1}-B_{a2}=0, \\
\mathcal{A}_{a}^{\prime 2} &\equiv &\pi _{a}^{2}+B_{a1}=0, \\
\mathcal{B}_{a}^{\prime \mu } &\equiv &\Pi _{a}^{\mu }=0.
\end{eqnarray}
The first Hamiltonian $\mathcal{H}^{\prime }$ is associated with the time
parameter $t \equiv x_{0}$. The Hamiltonians $\mathcal{A}_{a}^{\prime \mu}$ arose from the non-invertible
momenta $\pi _{a}^{\mu}$ and are related to the parameters $\lambda^a_\mu \equiv A_{\mu}^{a}$. Analogously, the Hamiltonians
$\mathcal{B}_{a}^{\prime \mu }$ are referred to the parameters $\epsilon _{\mu }^{a} \equiv B_{\mu }^{a}$.

The fundamental PB of the model are
\begin{eqnarray}
\left\{ A_{\mu }^{a}(x),\pi _{b}^{\nu }(x^{\prime })\right\}& =&\delta
_{b}^{a}\delta _{\mu }^{\nu }\delta ^{2}\left( \mathbf{x}-\mathbf{x}^{\prime }\right) ,\label{HJBF10}\\
\left\{ B_{\mu }^{a}(x),\Pi _{b}^{\nu }(x^{\prime })\right\} &=&\delta
_{b}^{a}\delta _{\mu }^{\nu }\delta ^{2}\left( \mathbf{x}-\mathbf{x}^{\prime }\right).\label{HJBF11}
\end{eqnarray}
The fundamental differential characterizes the evolution of any function of the phase space. It is expressed as
\begin{eqnarray}
df(x)=\int \left( \left\{ f(x),\mathcal{H}^{\prime }(x^{\prime })\right\}
dt+\left\{ f(x),\mathcal{A}_{a}^{\prime \mu }(x^{\prime })\right\} d\lambda
_{\mu }^{a}+\left\{ f(x),\mathcal{B}_{a}^{\prime \mu }(x^{\prime })\right\}
d\epsilon _{\mu }^{a}\right) d^{2}x^{\prime }.\label{HJBF12}
\end{eqnarray}
Now we check the integrability of the HJPDE. When the IC is applied to the Hamiltonians
$\mathcal{A}_{a}^{\prime1}$, $\mathcal{A}_{a}^{\prime 2}$, $\mathcal{B}_{a}^{\prime 1}$ and $%
\mathcal{B}_{a}^{\prime 2}$ we get relations of dependence between the
parameters related to them. This information tells us that these
Hamiltonians are non-involutive and can be used to construct the GB.

Let us rename $h_{a}^{0}\equiv \mathcal{A}_{a}^{\prime 1}$, $h_{a}^{1}\equiv \mathcal{A}_{a}^{\prime 2}$, $%
h_{a}^{2}\equiv \mathcal{B}_{a}^{\prime 1}$ and $h_{a}^{3}\equiv \mathcal{B}_{a}^{\prime 2}$. Let us denote $I,J=0,1,2,3$
as the indices of the elements of the matrix $M^{IJ}_{ab}(x,y) \equiv \{h^I_a(x),h^J_b(y)\}$. We have
\begin{equation*}
 M(x,y)=%
\begin{pmatrix}
0 & 0 & 0 & -1 \\
0 & 0 & 1 & 0 \\
0 & -1 & 0 & 0 \\
1 & 0 & 0 & 0%
\end{pmatrix}%
\delta ^{ab}\delta ^{2}(\mathbf{x}-\mathbf{x}^\prime).
\end{equation*}
This matrix has inverse
\begin{equation*}
M^{-1}(x,y)=%
\begin{pmatrix}
0 & 0 & 0 & 1 \\
0 & 0 & -1 & 0 \\
0 & 1 & 0 & 0 \\
-1 & 0 & 0 & 0%
\end{pmatrix}%
\delta ^{ab}\delta ^{2}(\mathbf{x}-\mathbf{x}^\prime ),
\end{equation*}
with this inverse we define the GB as
\begin{equation*}
\left\{ f(x),g(x^{\prime })\right\} ^{\ast }=\{f(x),g(x^{\prime })\}-\int
\{f(x),h_{c}^{I}(y)\}\left[ M^{-1}(y,y^{\prime })\right] _{IJ}^{cd}%
\{h_{d}^{J}(y^{\prime }),g(x^{\prime })\}dydy^{\prime }.\label{HJBF13}
\end{equation*}
We can use this expression to find the fundamental GB of the theory. The
non-vanishing results are given bellow:
\begin{eqnarray}
\left\{ A_{\mu }^{a}(x),\pi _{b}^{\nu }(x^{\prime })\right\} ^{\ast } &=& \delta
_{b}^{a}\delta _{\mu }^{\nu }\delta ^{2}(\mathbf{x}-\mathbf{x}^\prime ),\label{HJBF14}\\
\left\{ B_{0}^{a}(x),\Pi _{b}^{0}(x^{\prime })\right\} ^{\ast } &=& \delta
_{b}^{a}\delta ^{2}(\mathbf{x}-\mathbf{x}^\prime ),\label{HJBF15}\\
\left\{ A_{\mu }^{a}(x),B_{\nu }^{b}(x^{\prime })\right\} ^{\ast } &=& \delta
^{ab}\epsilon _{0\mu \nu }\delta ^{2}(\mathbf{x}-\mathbf{x}^\prime ).\label{HJBF16}
\end{eqnarray}
Comparing with the original PB (\ref{HJBF10},\ref{HJBF11}), we notice that $B^i_a, \Pi^a_i$ are no longer
conjugated variables. In fact, the $B^1_a$ now plays the role of $-\pi _{a}^{2}$ and $B^2_a$ the role of $\pi _{a}^{1}$. Only
$B_{0}^{a}(x),\Pi _{a}^{0}$ and $A_{\mu }^{a},\pi _{b}^{\nu }$ remains as conjugated variables.

After building the GB, the fundamental differential (\ref{HJBF12}) now takes the form
\begin{equation}
df(x)=\int \left( \left\{ f(x),\mathcal{H}^{\prime }(x^{\prime })\right\}
^{\ast }dt+\left\{ f(x),\mathcal{A}_{a}^{\prime 0}(x^{\prime })\right\}
^{\ast }d\lambda _{0}^{a}+\left\{ f(x),\mathcal{B}_{a}^{\prime 0}(x^{\prime
})\right\} ^{\ast }d\epsilon _{0}^{a}\right) d^{2}x^{\prime }.\label{HJBF17}
\end{equation}
We still need to analyse the IC of the Hamiltonians $\mathcal{A}_{a}^{\prime 0}$ and $\mathcal{B}%
_{a}^{\prime 0}$. By imposing $d\mathcal{A}_{a}^{\prime 0}=0$ and $d\mathcal{B}_{a}^{\prime 0}=0$ we
notice that we need to introduce two new Hamiltonians:
\begin{eqnarray}
\mathcal{C}^{\prime a} & \equiv & \epsilon ^{0\gamma \nu }D_{\gamma }B_{\nu
}^{a}=0, \label{HJBF18}\\
\mathcal{D}^{\prime a} & \equiv & \frac{1}{2}\epsilon ^{0\gamma \nu }\left[ F_{\gamma\nu}^{a}-\Lambda f_{\
bc}^{a} B_{\gamma }^{b}B_{\nu }^{c} \right] =0.\label{HJBF19}
\end{eqnarray}
Notice that the canonical Hamiltonian (\ref{HJBF07}) now can be written as $\mathcal H_0 = -A_{a0}\mathcal{C}^{\prime a}-B_{a0}\mathcal{D}^{\prime a}$.
The fields $\mathcal{A}_{a}^{\prime 0}$ and $\mathcal{B}_{a}^{\prime 0}$ have the role of Lagrange multipliers since they are coefficients of the constraints
in the canonical Hamiltonian. The new constrains also satisfy the IC and there is no need to
introduce new constrains or redefine the algebra. The integrability programme is then achieved and the complete set of involutive
Hamiltonians is  $\mathcal{A}_{a}^{\prime 0},\mathcal{B}_{a}^{\prime 0},\mathcal{C}^{\prime a},\mathcal{D}^{\prime a}$.

Let us define
\begin{eqnarray}
C^{\prime a}(\alpha) &\equiv& \int \alpha (y)\mathcal{C}^{\prime a}(y)d^{2}y,\label{HJBF20}\\
D^{\prime a}(\beta)  &\equiv& \int \beta (y)\mathcal{D}^{\prime a}(y)d^{2}y, \label{HJBF21}
\end{eqnarray}
where $\alpha $ and $\beta $ are weight functions. It follows the relation
\begin{eqnarray}
\left\{ C^{\prime a}(\alpha _{1}),C^{\prime b}(\alpha _{2})\right\} ^{\ast
}=f_{\rule{9pt}{0pt}c}^{ab}C^{\prime c}(\alpha _{1},\alpha _{2}),\label{HJBF22}\\
\left\{ C^{\prime a}(\alpha _{1}),D^{\prime b}(\beta _{1})\right\} ^{\ast
}=f_{\rule{9pt}{0pt}c}^{ab}D^{\prime c}(\alpha _{1},\beta _{1}),\label{HJBF23}\\
\left\{ D^{\prime a}(\beta _{1}),D^{\prime b}(\beta _{2})\right\} ^{\ast
}=-\Lambda f_{\rule{9pt}{0pt}c}^{ab}C^{\prime c}(\beta _{1},\beta _{2}).\label{HJBF24}
\end{eqnarray}
Note that for $\Lambda=0$, i.e., the pure three-dimensional gravity, the Hamiltonians satisfy the Poincaré algebra $ISO(2,1)$, we also
identify $D^{\prime a}$, which now commute with all the other Hamiltonians as the generator of translations.
For $\Lambda \neq 0$, the Hamiltonians close the $AdS$ or $dS$ algebra.

\section{Characteristic Equations of the 3D BF Gravity}

The IC allows us to find the complete set of involutive Hamiltonians: $\mathcal{A}_{a}^{\prime 0},\mathcal{B}_{a}^{\prime 0},\mathcal{C}^{\prime a},\mathcal{D}^{\prime a}$, all of them play a role in the evolution of the systems and must be added in the fundamental differential.
Let us rename
\begin{eqnarray*}
\mathcal{H}_{a}^{\prime 0} &\equiv &\mathcal{A}_{a}^{\prime 0}\quad
\longrightarrow \quad \omega _{a}^{0}, \\
\mathcal{H}_{a}^{\prime 1} &\equiv &\mathcal{B}_{a}^{\prime 0}\quad
\longrightarrow \quad \omega _{a}^{1}, \\
\mathcal{H}_{a}^{\prime 2} &\equiv &\mathcal{C}_{a}^{\prime }\quad
\longrightarrow \quad \omega _{a}^{2}, \\
\mathcal{H}_{a}^{\prime 3} &\equiv &\mathcal{D}_{a}^{\prime }\quad
\longrightarrow \quad \omega _{a}^{3},
\end{eqnarray*}
where the $\omega_a$ are the respective parameters. The final form of the fundamental differential is
\begin{equation}
df(x)=\int dx' \left( \left\{ f(x),\mathcal{H}^{\prime }(x^{\prime })\right\}
^{\ast }dt+\sum^3_{\kappa=0} \left\{ f(x),\mathcal{H}_{a}^{\prime \kappa}(x^{\prime })\right\}
^{\ast }d\omega _{a}^{\kappa}\right).\label{HJBF25}
\end{equation}
The CE are obtained from (\ref{HJBF25}), by evaluating $f$ for the fields $(A_{\mu }^{a},B_{\mu }^{a})$
and the momenta $(\pi^\mu_a,\Pi^\mu_a)$. For the first set we have
\begin{eqnarray}
dA_{\mu }^{a} &=&\delta _{\mu }^{0}\delta ^{ab}d\omega _{b}^{0}+ \delta^i_\mu \left[\left( D_i A_{0}^{a}-\Lambda f_{\rule{4pt}{0pt}%
bc}^{a}B_{i}^{b}B_{0}^{c}\right) dt- \delta^{ab}D _{i} d\omega^{2}_b+\Lambda f_{\rule%
{9pt}{0pt}c}^{ab}B_{i}^{c}d\omega _{b}^{3}\right],\label{HJBF26}\\
dB_{\mu }^{a} &=&\delta _{\mu }^{0}d\omega ^{a1}+
\delta^i_\mu \left[\left(D_{i}B_{0}^{a}-f_{\rule{4pt}{0pt}bc}^{a}A_{0}^{b}B_{i}^{c}\right)
dt-f_{\rule{9pt}{0pt}c}^{ab}B_{i}^{b}d\omega _{c}^{2}-\delta ^{ab}D_i d\omega _{b}^{3}\right],\label{HJBF27}
\end{eqnarray}
and
\begin{eqnarray}
d\pi _{a}^{\mu } &=&\epsilon ^{0\gamma \rho}  \left[ \delta _{0}^{\mu }
D_{\gamma }B_{a\rho }-\delta _{\rho }^{\mu }\left(
D_{\gamma }B_{a0}-f_{a}^{\rule{4pt}{0pt}bc}A_{b0}B_{c\gamma }\right) \right]
dt\nonumber\\
&+& \epsilon ^{0\gamma \rho}\left\{\delta _{\gamma}^{\mu }f_{a}^{\rule{4pt}{0pt}bc}B_{c\rho }d\omega _{b}^{2}
+\delta _{\gamma }^{\mu }\delta _{a}^{b}D _\gamma d\omega _{b}^{3}\right\},\label{HJBF28}\\
d\Pi _{a}^{\mu } & = &\delta _{0}^{\mu }\mathcal{H}_{a}^{\prime 3}dt.\label{HJBF29}
\end{eqnarray}
The integrability condition ensures the independence between the parameters related to the
involutive set of HJPDE. Therefore, since $t=x^0$ is one of these parameters, we can analyse
the temporal evolution of the fields independently. We have
\begin{eqnarray}
\partial_0 A_{\mu }^{a} &=& \delta^i_\mu \left( D_i A_{0}^{a}-\Lambda f_{\rule{4pt}{0pt}%
bc}^{a}B_{i}^{b}B_{0}^{c}\right) , \label{HJBF29a}\\
\partial_0 B_{\mu }^{a} &=& \delta^i_\mu \left(D_{i}B_{0}^{a}-f_{\rule{4pt}{0pt}bc}^{a}A_{0}^{b}B_{i}^{c}\right).\label{HJBF29b}
\end{eqnarray}
Note that the component $\mu=0$ of these equations states that $A_0^a,B_0^a$ are time independent parameters. This reinforce
the character of lagrange multipliers of these variables in the canonical Hamiltonian. On the other hand, the spatial components
of (\ref{HJBF29a}) are equivalent to the equation (\ref{HJBF02}). Analogously, the spatial components of (\ref{HJBF29b}) resemble equations
(\ref{HJBF03}).

For the second set of CE, we have
\begin{eqnarray}
\partial_0 \pi _{a}^{\mu } &=&\epsilon ^{0\gamma \rho}  \left[ \delta _{0}^{\mu }
D_{\gamma }B_{a\rho }-\delta _{\rho }^{\mu }\left(D_{\gamma }B_{a0}-f_{a}^{\ bc}A_{b0}B_{c\gamma }\right) \right], \label{HJBF29c}\\
\partial_0 \Pi _{a}^{\mu } & = &\delta _{0}^{\mu }\mathcal{H}_{a}^{\prime 3}.\label{HJBF29d}
\end{eqnarray}
Note that, the temporal evolution of the component $\pi_{a}^{\mu}$ is equal to the Hamiltonian $\mathcal{H}%
_{2}^{\prime a}=0$, leaving $\pi_{a}^{0}$ undetermined, just as its correspondent conjugated variable $A^a_0$.
For the component $\pi _{a}^{i}$, we have that its temporal evolution equation is in agreement with the definition of canonical momenta.
For $\Pi^{\mu}_a$, we have that its temporal evolution
is equal to zero. This result is in agreement with the fact that $\Pi^0_a$ is conjugated to a Lagrange multiplier and the $\Pi^i_a$ is no longer
a canonical variable.

\subsection{Generators of canonical and gauge transformations}

As it was shown in section $2$, the CE also give us the generator of the canonical transformations. In our case, we
need to consider the variations along the independent parameters $\omega _{a}$.
\begin{eqnarray}
dA_{\mu }^{a} &=& \delta _{\mu }^{0}\delta ^{ab}d\omega _{b}^{0} - \delta^i_\mu\delta^{ab}D _{i} d\omega^{2}_b - \delta^i_\mu\Lambda f_{\rule%
{9pt}{0pt}c}^{ab}B_{i}^{c}d\omega _{b}^{3},\label{HJBF30}\\
dB_{\mu }^{a} &=& \delta _{\mu }^{0}d\omega ^{a1} - \delta^i_\mu f_{\rule{9pt}{0pt}c}^{ab}B_{i}^{b}d\omega _{c}^{2} -
\delta^i_\mu \delta^{ab}D_i d\omega _{b}^{3}.\label{HJBF31}
\end{eqnarray}
These expressions can be rewritten in a much simple form if we define the
function
\begin{equation}
G^{can}\equiv \int \left[ \mathcal{H}_{0}^{\prime a}d \omega _{a}^{0}+%
\mathcal{H}_{1}^{\prime a}d \omega _{a}^{1}+\mathcal{H}_{2}^{\prime
a}d \omega _{a}^{2}+\mathcal{H}_{3}^{\prime a}d\omega _{a}^{3}%
\right] d^{2}x.\label{HJBF32}
\end{equation}
It enables us to write
\begin{eqnarray}
d A_{\mu }^{a}=\left\{ A_{\mu }^{a},G^{can}\right\}^{\ast },\\
d B_{\mu }^{a}=\left\{ B_{\mu }^{a},G^{can}\right\}^{\ast }.
\label{conj2}
\end{eqnarray}
As the variations of the phase space coordinates can be expressed in this
way, we call $G^{can}$ the \textit{generator of canonical transformations}.

On the other hand, in order to relate generator of canonical transformations with the one of symmetries, we need
to go further the IC. Let us consider the set of variations $(\ref{HJBF30})$ and $(\ref{HJBF31})$ now rewritten as
\begin{eqnarray}
\delta A_{\mu }^{a} &=& \delta _{\mu }^{0}\delta ^{ab}\delta\omega _{b}^{0} - \delta^i_\mu\delta^{ab}D _{i} \delta\omega^{2}_b - \delta^i_\mu\Lambda f_{\rule%
{9pt}{0pt}c}^{ab}B_{i}^{c}\delta\omega _{b}^{3},\label{HJBF33}\\
\delta B_{\mu }^{a} &=& \delta _{\mu }^{0}d\omega ^{a1} - \delta^i_\mu f_{\rule{9pt}{0pt}c}^{ab}B_{i}^{b}\delta\omega _{c}^{2} -
\delta^i_\mu\delta ^{ab}D_i \delta\omega _{b}^{3},\label{HJBF34}
\end{eqnarray}
where the variations $\delta\omega^\kappa_{a}$ may depend on each other. If the variations $(\ref{HJBF33})$,$(\ref{HJBF34})$ are symmetries of the
three-dimensional BF gravity, they must be solutions of the fixed point variation
\begin{eqnarray}
\delta \mathcal L = \frac{1}{2} \epsilon^{\alpha\mu\nu} \left( F^a_{\mu\nu}
- \Lambda f^a_{\ bc} B^b_\mu B^c_\nu \right) \delta B^a_\alpha + \epsilon^{\alpha\mu\nu} B^a_\mu D_\nu \delta A_{a\alpha} = 0.\label{HJBF35}
\end{eqnarray}
By replacing the $(\ref{HJBF33})$,$(\ref{HJBF34})$ in $(\ref{HJBF35})$ and using the Bianchi identity it follows that
\begin{eqnarray}
\delta \mathcal L &=& \epsilon^{ij}\left[\frac{1}{2}F_{ij}^{a}\left(\delta\omega^{a1}+f_{abc}B_{0}^{b}\delta\omega_{c}^{2}\right)
+B_{i}^{a}D_{j}\left(D_{0}\delta\omega_{a}^{2}+\delta\omega_{a}^{0}\right)+F_{a0j}D_{i}\delta\omega^{a3}\right]\nonumber\\
&+&	- \Lambda f_{abc}\epsilon^{ij}\left[\frac{1}{2}B_{i}^{b}B_{j}^{c}\left(\delta\omega^{a1}+f_{\ nm}^{a}B_{0}^{n}\delta\omega^{m2}\right)
-B_{0}^{a}B_{j}^{c}D_{i} \delta\omega^{b3}\right]\nonumber\\
&+&	- \Lambda f_{abc}\epsilon^{ij}\left[-B_{0}^{a}D_{i}\left(B_{j}^{b}\delta\omega^{c3}\right)+B_{i}^{a}D_{0}\left(B_{j}^{b}\delta\omega^{c3}\right)\right].
\label{HJBF36}
\end{eqnarray}
Since this is one equation for four parameters, we expect to obtain a relation between some of the $\delta \omega^\kappa_a$. A good approach to solve
$\delta \mathcal L = 0$ is by considering special cases, as setting some of the parameters equal to zero. However, by inspection of $(\ref{HJBF36})$, we see
that $\delta \omega^{a0}=0$ or $\delta \omega^{a1}=0$ are not good choices for solving the equation.
On the other hand, if we consider $\delta\omega^{a3}=0$, equation $(\ref{HJBF36})$ becomes
\begin{eqnarray}
\delta\mathcal{L} &=& \epsilon^{ij}\left[\frac{1}{2}F_{ij}^{a}\left(\delta\omega^{a1}+f_{abc}B_{0}^{b}\delta\omega_{c}^{2}\right)
+B_{i}^{a}D_{j}\left(D_{0}\delta\omega_{a}^{2}+\delta\omega_{a}^{0}\right)\right]\nonumber\\
&+&- \Lambda f_{abc}\epsilon^{ij}\left[\frac{1}{2}B_{i}^{b}B_{j}^{c}\left(\delta\omega^{a1}+f_{\ nm}^{a}B_{0}^{n}\delta\omega^{m2}\right)\right].\label{HJBF37}
\end{eqnarray}
Of course, we have an invariance, $\delta\mathcal{L}=0$, when we choose $\delta\omega_{a}^{0}=-D_{0}\delta\omega_{a}^{2}$ and
$\delta\omega^{a1}=	-f_{\ bc}^{a}B_{0}^{b}\delta\omega^{c2}$.  By replacing it in the set of HJ variations, we get
\begin{eqnarray}
\delta B_{\mu}^{a}	&=&	-f_{\ bc}^{a}B_{\mu}^{b}\delta\omega^{c2},\nonumber\\
\delta A_{a\mu}	&=&	-D_{\mu}\delta\omega_{a}^{2}.\nonumber
\end{eqnarray}
By setting $\omega^{a2}=-\chi^a$, we obtain the gauge transformation (\ref{3DBF03}). These transformations are generated by
\begin{equation}
G^{Gauge}\equiv \int \left[ \mathcal{H}_{0}^{\prime a}D_{0}
+\mathcal{H}_{1}^{\prime b}f^a_{\ bc}B_{0}^{c}-\mathcal{H}_{2}^{\prime
a} \right]\delta \chi_{a}^{2} d^{2}x.\label{HJBF38}
\end{equation}

Now, if we set $\delta\omega^{a2}=0$, we obtain
\begin{eqnarray}
\delta\mathcal{L}&=&\epsilon^{ij}\left[\frac{1}{2}F_{ij}^{a}\delta\omega^{a1}+B_{i}^{a}D_{j}\delta\omega_{a}^{0}+F_{a0j}D_{i}\delta\omega^{a3}\right]\nonumber\\
&+&- \Lambda f_{abc}\epsilon^{ij}\left[\frac{1}{2}B_{i}^{b}B_{j}^{c}\delta\omega^{a1}-B_{0}^{a}D_{i}\left(B_{j}^{b}\delta\omega^{c3}\right)
+B_{i}^{a}D_{0}\left(B_{j}^{b}\delta\omega^{c3}\right)-B_{0}^{a}B_{j}^{c}D_{i}\delta\omega^{b3}\right],\nonumber
\end{eqnarray}
which, up to boundary terms, becomes
\begin{eqnarray}
\delta\mathcal{L}&=&\epsilon^{ij}\left[-D_{j}B_{i}^{a}\left(\delta\omega_{a}^{1}-\Lambda f_{abc}B_{0}^{b}\delta\omega^{c3}\right)
+\frac{1}{2}F_{aij}\left(D_{0}\delta\omega^{a3}+\delta\omega^{a0}\right)\right]\nonumber\\
&+&	- \epsilon^{ij}\Lambda f_{abc}\left[\frac{1}{2}B_{i}^{a}B_{j}^{b}\left(D_{0}\delta\omega^{c3}+\delta\omega^{c0}\right)\right].\label{HJBF39}
\end{eqnarray}
This variation is equal to zero if we set $\delta\omega^{a0}=-D_{0}\delta\omega^{a3}$ and
$\delta\omega_{a}^{1}= \Lambda f_{abc}B_{0}^{b}\delta\omega^{c3}$. Under these condition
\begin{eqnarray}
\delta B_{\mu}^{a}&=&-D_{\mu}\delta\omega^{a3},\nonumber\\
\delta A_{a\mu}&=& \Lambda f_{abc}B_{\mu}^{b}\delta\omega^{c3},\nonumber
\end{eqnarray}
The shift transformation (\ref{3DBF04}) can be obtained just by setting $\eta^a=-\omega^{a3}$ in the previous relations. Its correspondent
generator is given by
\begin{equation}
G^{shift}\equiv \int \left[ \mathcal{H}_{0}^{\prime a}D_{0}
- \Lambda f^a_{\ bc}\mathcal{H}_{1}^{\prime b}B_{0}^{c}-\mathcal{H}_{3}^{\prime a}\right]\delta\eta_{a}^{3} d^{2}x.\label{HJBF40}
\end{equation}
Therefore, we have obtained the gauge and shift transformations as well as its respective generators with the use of the HJ formalism.

\section{Final Remarks}

We have used the Hamilton-Jacobi formalism to analyse the constraint structure of the three-dimensional BF gravity with a cosmological constant $\Lambda$.
This procedure consisted in finding the complete set of involutive Hamiltonians that generates the dynamical evolution of the system. We achieved this
using the Frobenius' Integrability Condition over the initial set of HJPDE. We noticed that there is a subgroup of Hamiltonians
($\mathcal A'^1_a, \mathcal A'^2_a, \mathcal B'^a_1, \mathcal B'^a_2$) that does not satisfy the IC and with them we built the GB and reduced the phase space such that the system was governed
by a new symplectic structure. By satisfying the IC for the rest of Hamiltonians, we found new constraints ($\mathcal C'_a, \mathcal D'_a$).
In the case of cosmological constant equal zero, these Hamiltonians satisfy the $ISO(1,2)$ algebra and the Hamiltonians
$\mathcal D'_a$ commute. When the cosmological constant is other than zero, the Hamiltonians satisfy the $AdS$ or $dS$ algebra.

Then, we computed the characteristic equations, which depend on the time parameter $x^0$ and the parameters $\omega^\kappa_a$ related to the involutive Hamiltonians.
Since all the Hamiltonians satisfy the IC, their correspondent parameters are linearly independent. It means that evolution along any parameter can be
considered independently. As a result, we saw that time evolution of the CE are equivalent to the field equations of BF gravity and the evolution along
the parameters $\omega^\kappa_a$ is related to the canonical transformations. Therefore, the linear combination of the four corresponding Hamiltonians gave
the generator of the canonical transformations.

It was possible to relate the generator of canonical transformations with the one related to the gauge and shift transformations. To achieve this, we considered
the $\omega^\kappa_a$ parameters as dependent on each other. Furthermore, if they are an invariance of the theory they must eliminate, up to boundary term, the
fixed point Lagrangian variation. This way, we needed to solve an equation for four dependent variables.

\bigskip

\section{Acknowledgements}

The authors thank M. C. Bertin for reading the manuscript and suggestions.
N. T. Maia was supported by CAPES. B. M. Pimentel was partially supported
by CNPq and CAPES. C. E. Valc\'{a}rcel was supported by FAPESP.

\end{document}